\documentclass[seceq,preprint]{ptptex}

\usepackage{amsmath}
\usepackage{amssymb}
\usepackage[dvipdfm]{graphicx}

\preprintnumber{IU-TH-12}

\title{
Reality and hermiticity from maximizing overlap 
in the future-included complex action theory

}

\author{%
Keiichi \textsc{Nagao}\footnote{E-mail: nagao@mx.ibaraki.ac.jp}
and Holger Bech \textsc{Nielsen}\footnote{E-mail: hbech@nbi.dk}
}

\inst{
$^{*)}$Faculty of Education, Ibaraki University, Mito 310-8512 
Japan \\
$^{**)}$Niels Bohr Institute, University of Copenhagen, 
17 Blegdamsvej Copenhagen $\phi$, Denmark
}

\abst{%

In the complex action theory whose path runs 
over not only past but also future 
we study 
a normalized matrix element of an operator $\hat{\cal O}$ 
defined in terms of the future state at the latest time $T_B$ and 
the past state at the earliest time $T_A$ with 
a proper inner product that makes normal 
a given Hamiltonian that is non-normal at first. 
We present a theorem that states that, 
provided that the operator $\hat{\cal O}$ is $Q$-Hermitian, i.e., 
Hermitian with regard to the proper inner product, 
the normalized matrix element becomes real 
and time-develops under 
a $Q$-Hermitian Hamiltonian 
for the past and future states selected such that 
the absolute value of the transition amplitude from 
the past state to the future state is maximized. 
Furthermore, we give a possible procedure 
to formulate the $Q$-Hermitian Hamiltonian 
in terms of $Q$-Hermitian coordinate and momentum operators, 
and construct a conserved probability current density. 
}

\begin{document}

\maketitle


\section{Introduction} 

The complex action theory (CAT) is a trial to extend 
quantum theories so that the action is complex 
at a fundamental level, but effectively looks real. 
So far, the CAT has been investigated 
with the expectation that the imaginary part of the action 
would give some falsifiable predictions\cite{Bled2006,Nielsen:2008cm,Nielsen:2007ak,Nielsen:2005ub},  
and various interesting suggestions have been made for the Higgs mass\cite{Nielsen:2007mj}, 
quantum mechanical philosophy\cite{newer1,Vaxjo2009,newer2}, 
some fine-tuning problems\cite{Nielsen2010qq,degenerate}, 
black holes\cite{Nielsen2009hq}, 
de Broglie-Bohm particles, and a cut-off in loop diagrams\cite{Bled2010B}. 
In the CAT, the Hamiltonian $\hat{H}$ is generically non-normal, 
so it is not contained in 
the class of PT-symmetric non-Hermitian Hamiltonians 
that have been intensively  studied\cite{Bender:1998ke,Bender:1998gh,Mostafazadeh_CPT_ip_2002,
Mostafazadeh_CPT_ip_2003}. 
In ref.\cite{Nagao:2010xu}, 
introducing what we call the proper inner product $I_Q$ 
so that the eigenstates of the Hamiltonian become orthogonal to 
each other with respect to it, 
we presented a mechanism to effectively 
obtain a Hamiltonian that is $Q$-Hermitian, i.e., 
Hermitian with respect to the proper inner product, 
after a long time development. 
In ref.\cite{Nagao:2011za}, we proposed 
a complex coordinate and momentum formalism by 
explicitly constructing non-Hermitian operators 
of complex coordinate $q$ and momentum $p$ 
and their eigenstates, so that we can deal with complex $q$ and 
$p$ properly. 
In general, the CAT could be classified into two theories: 
one is the future-not-included theory, i.e., the theory 
including only a past time as an integration interval of time, 
and the other one is the future-included theory\cite{Bled2006}, 
which includes not only a past time but also a future time. 
Using the complex coordinate and momentum 
formalism\cite{Nagao:2011za} in the Feynman path integral, 
we found that the momentum relation 
is given by the usual expression $p= m \dot{q}$, where 
$m$ is a complex mass, in the future-included theory\cite{Nagao:2011is}, and another expression 
$p=\left(m_R + \frac{m_I^2}{m_R}\right) \dot{q}$, 
where $m_R$ and $m_I$ are the real and imaginary parts of $m$, 
respectively,  
in the future-not-included theory\cite{Nagao:2013eda}.  
The future-included theory is described by using 
the future state $| B (T_B) \rangle$ at the final time $T_B$ 
and the past state $| A (T_A) \rangle$ at the initial time $T_A$. 
In refs.\cite{Nagao:2012mj,Nagao:2012ye}
we studied the normalized matrix element\footnote{In the real action theory (RAT), the normalized matrix element 
$\langle \hat{\cal O} \rangle^{BA}$ is called the weak value\cite{AAV}, and has been intensively studied. 
For details, see ref.\cite{review_wv} and references therein.} 
$\langle \hat{\cal O} \rangle^{BA} 
\equiv \frac{ \langle B(t) |  \hat{\cal O}  | A(t) \rangle }{ \langle B(t) | A(t) \rangle }$, 
where $t$ is an arbitrary time ($T_A \leq t \leq T_B$), 
in the future-included theory, and found that, 
if we regard $\langle \hat{\cal O} \rangle^{BA}$ 
as an expectation value in the future-included theory, then we 
obtain the Heisenberg equation, Ehrenfest's theorem, and a conserved probability current density. 
This suggests that $\langle \hat{\cal O} \rangle^{BA}$ is a 
strong candidate for an expectation value 
in the future-included theory.  

In this letter we study in the future-included CAT 
a slightly modified quantity 
$\langle \hat{\cal O} \rangle_Q^{BA} 
\equiv \frac{ \langle B(t) |_Q  \hat{\cal O}  | A(t) \rangle }{ \langle B(t) |_Q A(t) \rangle }$, 
where $\langle B(t)|_Q\equiv \langle B(t)|Q$, and $Q$ is a Hermitian operator\footnote{In the special case of the Hamiltonian $\hat{H}$ being 
Hermitian, $Q$ is just a unit operator, 
so $\langle \hat{\cal O} \rangle_Q^{BA}$ 
corresponds to $\langle \hat{\cal O} \rangle^{BA}$.} that is used 
to define the proper inner product $I_Q$.  
The choice of $\langle \hat{\cal O} \rangle_Q^{BA}$ or 
$\langle \hat{\cal O} \rangle^{BA}$ is 
only a matter of notation as to 
what the state symbol $\langle B(t)|$ shall precisely mean. 
On the other hand, 
the choice of the inner product used in the normalization of the initial and final states $|A(T_A) \rangle$ and $\langle B(T_B)|$ 
is not just a matter of notation, once we have chosen 
$\langle \hat{\cal O} \rangle_Q^{BA}$ 
as the expression of the candidate for our expectation value. 
That is to say, according to the choice of the inner product 
used in the normalization of 
the initial and final states, two slightly different versions 
could be defined. 
The normalization defined with the usual inner product 
$I$ has the true meaning of normalization, of course, 
but includes unphysical transitions between different 
eigenstates with different eigenvalues of the Hamiltonian $\hat{H}$. 
The normalization defined with the proper inner product $I_Q$, 
which we call $Q$-normalization, excludes such unphysical transitions, 
but does not have the original meaning of normalization.  
Thus, each choice seems to have both advantages and disadvantages, 
so we are interested in the study of both versions. 
However, let us admit that, in the version with the usually normalized 
initial and final states, it is not easy to evaluate 
$\langle \hat{\cal O} \rangle_Q^{BA}$ clearly, because 
we cannot exhaustively make use of the orthogonality 
of the eigenstates of the Hamiltonian $\hat{H}$. 
Therefore, we postpone the study of this version to the future, 
and concentrate in this letter on the analysis of the version 
with the $Q$-normalized initial and final states, 
which is much easier to study than the other version.

%
%
Assuming that a given Hamiltonian $\hat{H}$ 
is non-normal but diagonalizable, and that 
the imaginary parts of the eigenvalues 
of $\hat{H}$ are bounded from above, 
we present a theorem that claims that 
$\langle \hat{\cal O} \rangle_Q^{BA}$ becomes real 
and time-develops under a $Q$-Hermitian Hamiltonian 
for any $Q$-Hermitian operator $\hat{\cal O}$, 
provided that $| B (t) \rangle$ and $| A (t) \rangle$ 
are the time-developed states 
maximizing the absolute value of the transition amplitude 
$|\langle B(t) |_Q A(t) \rangle|$. 
Such states would represent an approximation to 
$|\langle B(t) |_Q A(t) \rangle|$ 
in the situation that $| B (T_B) \rangle$ and $| A (T_A) \rangle$ 
were randomly given.
In fact, in the large $T\equiv T_B-T_A$ case, 
only terms associated with the largest imaginary parts of the eigenvalues of the Hamiltonian would dominate, 
and even with random initial and final states 
the dominant term would give the biggest value. 
We call this thinking the maximization principle. 
We shall prove this theorem by finding that 
$\langle \hat{\cal O} \rangle_Q^{BA}$ 
for the states maximizing $|\langle B(t) |_Q A(t) \rangle|$ 
becomes an expression similar to 
an expectation value defined with $I_Q$ in the future-not-included 
theory. 
Indeed, it is very important to obtain a real expectation value 
and a Hermitian Hamiltonian in the CAT so that it can 
survive as a possible true fundamental quantum theory. 
The maximization principle 
is regarded as a method of obtaining not only a real expectation 
value but also a $Q$-Hermitian Hamiltonian. 
Furthermore, assuming that the non-normal Hamiltonian given 
at first is written in terms of the Hermitian 
coordinate and momentum operators $\hat{q}$ and $\hat{p}$, 
we give a possible procedure to formulate 
the $Q$-Hermitian Hamiltonian 
in terms of $Q$-Hermitian coordinate and momentum 
operators $\hat{q}_Q$ and $\hat{p}_Q$. 
We also provide a $Q$-Hermitian 
probability density operator 
and construct a conserved probability current density.

\section{Proper inner product and future-included complex action theory}

We consider a general non-normal diagonalizable 
Hamiltonian $\hat{H}$, i.e., $[\hat{H}, \hat{H}^{\dagger}]\neq 0$, 
for a general quantum mechanical system that 
could be the whole world, and review a 
proper inner product for $\hat{H}$ that makes $\hat{H}$ 
normal with respect to it by following refs.\cite{Nagao:2010xu,Nagao:2011za}. 
We define the eigenstates
 $| \lambda_i \rangle (i=1,2,\dots)$ of $\hat{H}$ such that 
\begin{equation}
\hat{H} | \lambda_i \rangle = \lambda_i | \lambda_i \rangle, 
\end{equation}
where $\lambda_i (i=1,2,\dots)$ are the eigenvalues of 
$\hat{H}$, and introduce the diagonalizing operator 
$P=(| \lambda_1 \rangle , | \lambda_2 \rangle , \ldots)$, 
so that $\hat{H}$ is diagonalized as 
$\hat{H} = PD P^{-1}$, 
where $D$ is given by $\text{diag}(\lambda_1, \lambda_2, \dots)$. 
Let us consider a transition from an eigenstate 
$| \lambda_i \rangle$ to another 
$| \lambda_j \rangle ~(i \neq j)$ fast in time $\Delta t$.  
Since $| \lambda_i \rangle$ are not orthogonal to 
each other in the usual inner product $I$, 
$I(| \lambda_i \rangle , | \lambda_j \rangle ) \equiv \langle \lambda_i | \lambda_j \rangle \neq \delta_{ij}$, 
the transition can be measured, i.e., 
$|I(| \lambda_j \rangle, \exp\left( -\frac{i}{\hbar} \hat{H} \Delta t \right) |\lambda_i \rangle )|^2 \neq 0$, 
though $\hat{H}$ cannot bring the system from $| \lambda_i \rangle$ 
to $| \lambda_j \rangle ~(i \neq j)$. 
Such an unphysical transition from one eigenstate to another 
with a different eigenvalue should be prohibited in a reasonable theory. 
In order to have reasonable probabilistic results, 
we introduce 
a proper inner product \cite{Nagao:2010xu, Nagao:2011za}\footnote{Similar inner products are also studied 
in refs.\cite{Geyer,Mostafazadeh_CPT_ip_2002,Mostafazadeh_CPT_ip_2003}.} 
for arbitrary kets $|u \rangle$ and $|v \rangle$ as 
\begin{equation}
I_Q(|u \rangle , |v \rangle) \equiv \langle u |_Q v \rangle 
\equiv \langle u | Q | v \rangle,
\end{equation}  
where $Q$ is a Hermitian operator chosen as 
$Q=(P^\dag)^{-1} P^{-1}$, 
so that $| \lambda_i \rangle$ become orthogonal to each other 
with regard to $I_Q$: 
\begin{equation} 
\langle \lambda_i  |_Q \lambda_j \rangle = \delta_{ij}. 
\end{equation}
This implies the orthogonality relation 
$\sum_i | \lambda_i \rangle \langle \lambda_i |_{Q} = 1$. 
In the special case of the Hamiltonian $\hat{H}$ being 
Hermitian, $Q$ would be the unit operator. 
We introduce the ``$Q$-Hermitian'' conjugate $\dag^Q$ of 
an operator $A$ by 
$\langle \psi_2 |_Q A | \psi_1 \rangle^* \equiv \langle \psi_1 |_Q A^{\dag^Q} | \psi_2 \rangle$, so 
\begin{equation} 
A^{\dag^Q} \equiv Q^{-1} A^\dag Q. 
\end{equation}
If $A$ obeys $A^{\dag^Q} = A$, $A$ is $Q$-Hermitian. 
We also define $\dag^Q$ for kets and bras as 
$| \lambda \rangle^{\dag^Q} \equiv \langle \lambda |_Q $ and 
$\left(\langle \lambda |_Q \right)^{\dag^Q} \equiv | \lambda \rangle$. 
In addition, 
$P^{-1}=
\left(
 \begin{array}{c}
      \langle \lambda_1 |_Q     \\
      \langle \lambda_2 |_Q     \\
      \vdots 
 \end{array}
\right)$ 
satisfies $P^{-1} \hat{H} P = D$ and 
$P^{-1} \hat{H}^{\dag^Q} P = D^{\dag}$, 
so $\hat{H}$ is ``$Q$-normal'', 
$[\hat{H}, \hat{H}^{\dag^Q} ] = P [D, D^\dag ] P^{-1} =0$.  
Thus the inner product $I_Q$ makes $\hat{H}$ 
$Q$-normal. 
We note that $\hat{H}$ can be decomposed as 
$\hat{H}=\hat{H}_{Qh} + \hat{H}_{Qa}$, 
where $\hat{H}_{Qh}= \frac{\hat{H} + \hat{H}^{\dag^Q} }{2}$ and 
$\hat{H}_{Qa} = \frac{\hat{H} - \hat{H}^{\dag^Q} }{2}$ are 
$Q$-Hermitian and anti-$Q$-Hermitian parts of $\hat{H}$,  respectively.

In refs.\cite{Bled2006,Nagao:2012mj,Nagao:2012ye} 
the future-included theory is 
described by using 
the future state $| B (T_B) \rangle$ at the final time $T_B$ 
and the past state $| A (T_A) \rangle$ at the initial time $T_A$, 
where $| A (T_A) \rangle$ and $| B (T_B) \rangle$ 
time-develop as follows: 
\begin{eqnarray}
&&i \hbar \frac{d}{dt} | A(t) \rangle = \hat{H} | A(t) \rangle , \label{schro_eq_Astate} \\
&&-i \hbar \frac{d}{dt} \langle B(t) |  
= \langle B(t) |  \hat{H} , \label{schro_eq_Bstate_old} 
\end{eqnarray}
and the ``normalized'' matrix element 
$\langle \hat{\cal O} \rangle^{BA} 
\equiv \frac{ \langle B(t) |  \hat{\cal O}  | A(t) \rangle }{ \langle B(t) | A(t) \rangle }$ 
is studied. 
The quantity $\langle \hat{\cal O} \rangle^{BA}$ 
is called the weak value\cite{AAV,review_wv} 
in the real action theory (RAT).  
In refs.\cite{Nagao:2012mj,Nagao:2012ye} 
we investigated $\langle \hat{\cal O} \rangle^{BA}$, and found that, 
if we regard $\langle \hat{\cal O} \rangle^{BA}$ 
as an expectation value in the future-included theory, 
then we obtain the Heisenberg equation, Ehrenfest's theorem, 
and a conserved probability current density. 
Thus $\langle \hat{\cal O} \rangle^{BA}$ seems to play the 
role of an expectation value in the future-included theory.  

In this letter, we adopt the proper inner product $I_Q$ 
for all quantities, 
and hence slightly modify the final state 
$\langle B(T_B) |$ as 
$\langle B(T_B) | \rightarrow \langle B(T_B) |_Q$ 
so that the Hermitian operator $Q$ pops out 
and the usual inner product $I$ is replaced with $I_Q$. 
Our new final state $\langle B(T_B) |$ time-develops 
according not to 
eq.(\ref{schro_eq_Bstate_old}) but to 
\begin{eqnarray}
-i \hbar \frac{d}{dt} \langle B(t) |_Q  
= \langle B(t) |_Q  \hat{H} 
\quad \Leftrightarrow \quad 
i \hbar \frac{d}{dt} | B(t) \rangle = {\hat{H}}^{\dag^Q} | B(t) \rangle . 
\label{schro_eq_Bstate} 
\end{eqnarray}
Thus the normalized matrix element 
$\langle \hat{\cal O} \rangle^{BA}$ is modified into 
the following expression: 
\begin{equation}
\langle \hat{\cal O} \rangle_Q^{BA} 
\equiv \frac{ \langle B(t) |_Q  \hat{\cal O}  | A(t) \rangle }{ \langle B(t) |_Q A(t) \rangle }, 
\end{equation} 
where $I_Q$ is used for both the denominator and numerator.  
As far as the construction of $\langle \hat{\cal O} \rangle_Q^{BA}$ 
is concerned, 
the shift between $\langle B(t)|$ and $\langle B(t)|_Q$ is 
just a change of notation, but when it comes to our
maximization principle, we need to normalize 
the initial and final states $|A(T_A) \rangle$ and $\langle B(T_B)|$. 
There are two choices: 
the normalization 
defined with the usual inner product $I$ 
or 
the normalization defined with 
the proper inner product $I_Q$,  
which we call $Q$-normalization. 
The choice of the inner product used in the normalization  
is not just a matter of notation, 
once we have chosen 
$\langle \hat{\cal O} \rangle_Q^{BA}$ 
as the expression of the candidate for our expectation value. 
That is to say, according to the choice of the inner product 
used in the normalization of 
the initial and final states, two slightly different versions 
could be defined. 
As we have explained in the introduction, 
each choice seems to have both advantages and disadvantages, 
and it is not easy to evaluate 
$\langle \hat{\cal O} \rangle_Q^{BA}$ 
clearly in the version with the usually normalized 
initial and final states, because 
we cannot exhaustively make use of the orthogonality 
of the eigenstates of the Hamiltonian $\hat{H}$. 
Therefore, we postpone the study of this version to the future, and in the following 
we investigate 
the quantity $\langle \hat{\cal O} \rangle_Q^{BA}$ 
with the $Q$-normalized initial and final states $|A(T_A) \rangle$ and $\langle B(T_B)|$, 
which is much easier to study than the other version.

\section{Theorem on the normalized matrix element and its proof}

We present the following theorem:

{\bf Theorem} 
{\em 
As a prerequisite, assume that a given Hamiltonian 
$\hat{H}$ is non-normal but diagonalizable 
and that the imaginary parts of the eigenvalues 
of $\hat{H}$ are bounded from above, 
and define a modified inner product $I_Q$ by means 
of a Hermitian operator $Q$ arranged so 
that $\hat{H}$ becomes normal with respect to $I_Q$. 
Let the two states $| A(t) \rangle$ and $ | B(t) \rangle$ 
time-develop according to the Schr\"{o}dinger 
equations\footnote{See eqs.(\ref{schro_eq_Astate}) and (\ref{schro_eq_Bstate}).} 
with $\hat{H}$ and $\hat{H}^{\dag^Q}$, respectively: 
$|A (t) \rangle = 
e^{-\frac{i}{\hbar}\hat{H} (t-T_A) }| A(T_A) \rangle$, 
$|B (t) \rangle = 
e^{-\frac{i}{\hbar} {\hat{H}}^{\dag^Q} (t-T_B) } 
| B(T_B)\rangle$, 
and be normalized with $I_Q$ 
at the initial time $T_A$ and the final time $T_B$, respectively: 
$\langle A(T_A) |_{Q} A(T_A) \rangle = 1$,
$\langle B(T_B) |_{Q} B(T_B) \rangle = 1$. 
Next determine
$|A(T_A) \rangle$ 
and $|B(T_B) \rangle$ so as to maximize 
the absolute value of the transition 
amplitude $|\langle B(t) |_Q A(t) \rangle|=
|\langle B(T_B)|_Q \exp(-i\hat{H}(T_B-T_A))
|A(T_A) \rangle|$. 
Then, provided that an operator $\hat{\cal O}$
is $Q$-Hermitian, i.e., Hermitian with respect to 
the inner product $I_Q$, 
$\hat{\cal O}^{\dag^Q} = \hat{\cal O}$, 
the normalized matrix element of 
the operator $\hat{\cal O}$ defined by 
$\langle \hat{\cal O} \rangle_Q^{BA} 
\equiv
\frac{\langle B(t) |_Q \hat{\cal O} | A(t) \rangle}{\langle B(t) |_Q A(t) \rangle}$ 
becomes {\rm real} 
and time-develops under 
a {\rm $Q$-Hermitian} Hamiltonian. }
%


\vspace*{0.5cm}

Before proving the theorem, 
we make a couple of remarks on it. 
The procedure of maximizing 
the absolute 
value of the transition amplitude 
$|\langle B(t) |_Q A(t) \rangle|$, 
which we call the maximization principle, 
can be understood as an approximation 
to what will be with very large likelihood 
the result of just taking the initial state $| A(T_A) \rangle$ 
and the final state $| B(T_B) \rangle$ 
at random. 
In fact, we would like to show 
in a later publication that with the random states 
$| A(T_A) \rangle$ and $| B(T_B) \rangle$ 
we obtain approximately 
the same result for $\langle \hat{\cal O} \rangle_Q^{BA}$ 
as if we used the 
maximization principle as just stated in the theorem. 
The crucial point of the theorem is that 
$\langle \hat{\cal O} \rangle_Q^{BA}$, which is taken as an average 
for an operator $\hat{\cal O}$ obeying $\hat{\cal O}^{\dag^Q} =\hat{\cal O}$, 
turns out to be real almost unavoidably. 
This is under the restriction that $\hat{H}$ be $Q$-normal, 
i.e., normal with regard to the proper inner product $I_Q$, 
but that $\hat{H}$ is not required to be $Q$-Hermitian, 
$\hat{H} \neq \hat{H}^{\dag^Q}$.

Now let us prove the above theorem 
by expanding $| A(t) \rangle$ and $| B(t) \rangle$ 
in terms of the eigenstates $| \lambda_i \rangle$ 
as follows: 
$|A (t) \rangle = \sum_i a_i (t) | \lambda_i \rangle$, 
$|B (t) \rangle = \sum_i b_i (t) | \lambda_i \rangle$,  
where 
$a_i (t) = a_i (T_A) e^{-\frac{i}{\hbar}\lambda_i (t-T_A) }$, 
$b_i (t) = b_i (T_B) e^{-\frac{i}{\hbar}\lambda_i^* (t-T_B) }$. 
Since $\langle B (t) |_Q A (t) \rangle$ is expressed as 
$\langle B (t) |_Q A (t) \rangle 
= \sum_i R_i e^{i \Theta_i}$, 
where we have introduced 
$a_i(T_A)= | a_i(T_A) | e^{i \theta_{a_i}}$, 
$b_i(T_B) = | b_i(T_B) | e^{i \theta_{b_i}}$, 
$T\equiv T_B - T_A$, 
$R_i
\equiv |a_i (T_A)| |b_i (T_B)| e^{\frac{1}{\hbar} T \text{Im} \lambda_i }$, 
and $\Theta_i \equiv \theta_{a_i} - \theta_{b_i} 
- \frac{1}{\hbar} T \text{Re} \lambda_i$, 
$| \langle B (t) |_Q A (t) \rangle |^2$ is calculated as 
$| \langle B (t) |_Q A (t) \rangle |^2
= \sum_i R_i^2 + 2 \sum_{i<j} R_i R_j \cos(\Theta_i - \Theta_j)$. 
On the other hand, the normalization conditions  
are expressed as 
$\sum_i  | a_i (T_A) |^2  = 1$ and 
$\sum_i  | b_i (T_B) |^2  = 1$, respectively.

Here we note that 
the imaginary parts of the eigenvalues 
of $\hat{H}$ have to be bounded from above 
to avoid the Feynman path integral
$\int e^{\frac{i}{\hbar}S} {\cal D} \text{path}$ 
being divergently meaningless. 
So we assume that some of the $\text{Im} \lambda_i$ 
take the maximal value $B$, and denote 
the corresponding subset of $\{ i \}$ as $A$. 
Then, since $R_i \geq 0$, 
$| \langle B (t) |_Q A (t) \rangle |$ can take 
a maximal value only under the following conditions: 
\begin{eqnarray}
&& | a_i (T_A) |  = | b_i (T_B) | =0 \quad \text{for $\forall i \notin A$} , \label{abinotinA0} \\
&& \Theta_i  
\equiv \Theta_c 
\quad \text{for $\forall i \in A$} \label{max_cond_theta} , \\ 
&& \sum_{i \in A} | a_i (T_A) |^2 =\sum_{i \in A}|b_i (T_B)|^2 = 1,  \label{nc_ATABTB3} 
\end{eqnarray}
and $| \langle B (t) |_Q A (t) \rangle |^2$ is estimated as 
\begin{eqnarray}
| \langle B (t) |_Q A (t) \rangle |^2
&=& \left( \sum_{i \in A} R_i \right)^2  \nonumber \\  
&=& e^{\frac{2 B T}{\hbar} } 
\left( \sum_{i \in A} |a_i (T_A)| |b_i (T_B)| \right)^2   \nonumber \\
&\leq& e^{\frac{2 B T}{\hbar} } 
\left\{ \sum_{i \in A} \left( \frac{ |a_i (T_A)| + |b_i (T_B)|}{2} \right)^2 \right\}^2 
=e^{\frac{2}{\hbar} B T} , 
\end{eqnarray}
where the third equality is realized for 
\begin{equation}
 |a_i (T_A)| = |b_i (T_B)|  \quad \text{for $\forall i \in A$}. 
\label{max_cond_ab}  
\end{equation}
In the last equality we have used this relation 
and eq.(\ref{nc_ATABTB3}). 
The maximization condition of  $| \langle B (t) |_Q A (t) \rangle |$ 
is represented by 
eqs.(\ref{abinotinA0})-(\ref{nc_ATABTB3}) and (\ref{max_cond_ab}). 
That is to say, 
the states to maximize $| \langle B (t) |_Q A (t) \rangle |$, 
$| A(t) \rangle_{\rm{max}}$ and $| B(t) \rangle_{\rm{max}}$, are 
expressed as 
\begin{eqnarray}
&&|A (t) \rangle_{\rm{max}} = \sum_{i \in A} a_i (t) | \lambda_i \rangle , 
\label{Atketmax_sum_inA_ai} \\
&&|B (t) \rangle_{\rm{max}} = \sum_{i \in A} b_i (t) | \lambda_i \rangle, 
\label{Btketmax_sum_inA_bi} 
\end{eqnarray}
where $a_i (t)$ and $b_i (t)$ obey 
eqs.(\ref{max_cond_theta}), (\ref{nc_ATABTB3}), and (\ref{max_cond_ab}). 
Intuitively, it might be rather obvious that, to 
get the biggest transition amplitude 
$| \langle B (t) |_Q A (t) \rangle |$ for states 
$| A (t) \rangle $ and $| B(t) \rangle$ normalized 
at the initial time $T_A$ and the final time 
$T_B$, respectively, we should 
seek the eigenstates leading to the 
biggest increase with time development under the Schr\"{o}dinger equations, i.e., with the biggest 
imaginary parts of the eigenvalues of $\hat{H}$.

We evaluate $\langle \hat{\cal O} \rangle_Q^{BA}$ 
for $| A(t) \rangle_{\rm{max}} $ and $| B(t) \rangle_{\rm{max}}$. 
Using eqs.(\ref{abinotinA0})-(\ref{nc_ATABTB3}) and (\ref{max_cond_ab}), 
we obtain 
${}_{\rm{max}} \langle B (t) |_Q A (t) \rangle_{\rm{max}} 
= e^{i \Theta_c} \sum_{i \in A} R_i  
= e^{i \Theta_c} e^{\frac{B T}{\hbar} }$, 
and 
\begin{eqnarray}
{}_{\rm{max}} \langle B (t) |_Q \hat{\cal O} | A  (t) \rangle_{\rm{max}} 
&=& 
e^{i \Theta_c} e^{\frac{B T}{\hbar} }
\sum_{i , j \in A} a_j(T_A)^*  a_i(T_A) 
e^{\frac{i}{\hbar}(t-T_A) (\text{Re}\lambda_j - \text{Re}\lambda_i)} 
\langle \lambda_j |_Q \hat{\cal O} | \lambda_i \rangle  \nonumber \\
&=& 
e^{i \Theta_c} e^{\frac{B T}{\hbar} }
\langle \tilde{A}(t) |_Q \hat{\cal O} | \tilde{A}(t) \rangle , 
\end{eqnarray}
where we have introduced 
$| \tilde{A}(t) \rangle \equiv 
e^{-\frac{i}{\hbar}(t-T_A) \hat{H}_{Qh}} | A(T_A) \rangle_{\rm{max}}$, 
which is normalized as 
$\langle \tilde{A}(t) |_Q \tilde{A}(t) \rangle = 1$ 
and obeys the Schr\"{o}dinger equation 
\begin{eqnarray}
i\hbar  \frac{d}{d t}| \tilde{A}(t) \rangle 
&=& \hat{H}_{Qh} | \tilde{A}(t) \rangle .  \label{ScheqAtildetket}
\end{eqnarray} 
Thus the normalized matrix element 
$\langle \hat{\cal O} \rangle_Q^{BA}$  
for $| A(t) \rangle_{\rm{max}}$ and $| B(t) \rangle_{\rm{max}}$ 
is evaluated as 
\begin{eqnarray}
\langle \hat{\cal O} \rangle_Q^{BA} 
&=& 
\langle \tilde{A}(t) |_Q \hat{\cal O} | \tilde{A}(t) \rangle 
\equiv 
\langle \hat{\cal O} \rangle_Q^{\tilde{A} \tilde{A}} . \label{OBAmaxtilde}
\end{eqnarray}
Now we see that $\langle \hat{\cal O} \rangle_Q^{BA}$ 
for $| A(t) \rangle_{\rm{max}} $ and $| B(t) \rangle_{\rm{max}}$
has become the form of an average 
defined with the proper inner product $I_Q$. 
Since the complex conjugate of 
$\langle \hat{\cal O} \rangle_Q^{\tilde{A} \tilde{A}}$ 
is expressed as 
$\left\{ \langle \hat{\cal O} \rangle_Q^{\tilde{A} \tilde{A}} \right\}^*=\langle \hat{\cal O}^{\dag^Q} \rangle_Q^{\tilde{A} \tilde{A}}$, 
$\langle \hat{\cal O} \rangle_Q^{BA}$ 
for $| A(t) \rangle_{\rm{max}} $ and $| B(t) \rangle_{\rm{max}}$ 
is shown to be real for $Q$-Hermitian $\hat{\cal O}$.

Next we study the time development of 
$\langle \hat{\cal O} \rangle_Q^{\tilde{A} \tilde{A}}$. 
We express $\langle \hat{\cal O} \rangle_Q^{\tilde{A} \tilde{A}}$ as 
$\langle \hat{\cal O} \rangle_Q^{\tilde{A} \tilde{A}}
=\langle \tilde{A}(T_A) |_Q$ $\hat{\cal O}_{H}(t, T_A)
 | \tilde{A}(T_A) \rangle$, 
where we have introduced the Heisenberg operator 
$\hat{\cal O}_{H}(t, T_A) 
\equiv 
e^{ \frac{i}{\hbar} \hat{H}_{Qh} (t-T_A) }$ 
$\hat{\cal O} 
e^{ -\frac{i}{\hbar} \hat{H}_{Qh} (t-T_A)}$.  
This operator $\hat{\cal O}_{H}(t, T_A)$ obeys 
the Heisenberg equation 
\begin{equation}
i\hbar  \frac{d}{d t} \hat{\cal O}_{H}(t, T_A) 
= [ \hat{\cal O}_{H}(t, T_A) , \hat{H}_{Qh} ], 
\end{equation}
so we find that 
$\langle \hat{\cal O} \rangle_Q^{\tilde{A} \tilde{A}}$ time-develops 
under the $Q$-Hermitian Hamiltonian $\hat{H}_{Qh}$ as 
\begin{eqnarray}
\frac{d}{dt} \langle \hat{\cal O} \rangle_Q^{\tilde{A} \tilde{A}} 
&=&
\frac{i}{\hbar} 
\langle \left[ \hat{H}_{Qh}, \hat{\cal O} \right]  
\rangle_Q^{\tilde{A} \tilde{A}} . 
\label{ddtOAtildeAtildeQ}
\end{eqnarray}
Now, for pedagogical reasons, let us suppose that  
$\langle \hat{\cal O} \rangle_Q^{\tilde{A} \tilde{A}}$ time-develops 
under some Hamiltonian $\hat{H}_{1}$ as 
$\frac{d}{dt} \langle \hat{\cal O} \rangle_Q^{\tilde{A} \tilde{A}} 
=\frac{i}{\hbar} 
\langle \left[ \hat{H}_{1}, \hat{\cal O} \right]  \rangle_Q^{\tilde{A} \tilde{A}}$. 
The complex conjugate of this relation is given by 
$\left\{ \frac{d}{dt} \langle \hat{\cal O} \rangle_Q^{\tilde{A} \tilde{A}} \right\}^* 
=\frac{i}{\hbar} 
\langle \left[ \hat{H}_{1}^{\dag^Q}, \hat{\cal O}^{\dag^Q} \right]  \rangle_Q^{\tilde{A} \tilde{A}}$. 
Since $\langle \hat{\cal O} \rangle_Q^{\tilde{A} \tilde{A}}$ 
is real for $Q$-Hermitian $\hat{\cal O}$, 
these relations 
claim that $\hat{H}_{1}$ has to be $Q$-Hermitian. 
Therefore, 
the reality of $\langle \hat{\cal O} \rangle_Q^{\tilde{A} \tilde{A}}$ 
implies that it has to time-develop 
under some $Q$-Hermitian Hamiltonian. 
As shown in eq.(\ref{ddtOAtildeAtildeQ}), 
$\langle \hat{\cal O} \rangle_Q^{\tilde{A} \tilde{A}}$ time-develops 
under $\hat{H}_{Qh}$, which is consistent with the implication. 
We emphasize that 
the maximization principle 
provides not only 
the reality of $\langle \hat{\cal O} \rangle_Q^{BA}$ 
for $Q$-Hermitian $\hat{\cal O}$ 
but also the $Q$-Hermitian Hamiltonian.

\section{Discussion}

In this letter, 
we first reviewed the proper inner product $I_Q$ 
defined with a Hermitian operator $Q$, which 
is constructed from a diagonalizing operator of 
a given non-normal diagonalizable Hamiltonian $\hat{H}$, 
so that the eigenstates of $\hat{H}$ 
become orthogonal to each other with regard to 
the proper inner product $I_Q$, 
and the $Q$-Hermitian conjugate $\dag^Q$, i.e., 
Hermitian conjugate with regard to $I_Q$. 
We also explained the property of 
the normalized matrix element $\langle \hat{\cal O} \rangle^{BA}=\frac{\langle B(t) | \hat{\cal O} | A(t) \rangle}{\langle B(t) | A(t) \rangle}$ 
in the future-included complex action theory (CAT).  
Next we introduced a slightly modified  
normalized matrix element 
$\langle \hat{\cal O} \rangle_Q^{BA}=\frac{\langle B(t) |_Q \hat{\cal O} | A(t) \rangle}{\langle B(t) |_Q A(t) \rangle}$, 
which is defined with $I_Q$, 
and explained that two versions could be defined 
according to the choice of the normalization of 
the initial and final states $|A(T_A)\rangle$ and $\langle B(T_B)|$. 
One is the usual normalization defined with the usual inner 
product $I$, 
and the other is the $Q$-normalization defined with 
the proper inner product $I_Q$. 
Assuming that a given Hamiltonian $\hat{H}$ is 
non-normal but diagonalizable, and that  
the imaginary parts of the eigenvalues of $\hat{H}$ 
are bounded from above, 
we presented a theorem 
that states that, provided that $\hat{\cal O}$ is $Q$-Hermitian, i.e., 
$\hat{\cal O}^{\dag^Q}=\hat{\cal O}$, 
and that $|A(t) \rangle $ and $|B(t) \rangle$ time-develop 
according to the Schr\"{o}dinger equations 
with $\hat{H}$ and $\hat{H}^{\dag^Q}$ and are 
$Q$-normalized 
at the initial time $T_A$ and at the final time $T_B$, respectively, 
$\langle \hat{\cal O} \rangle_Q^{BA}$ becomes real 
and time-develops under a $Q$-Hermitian Hamiltonian  
for $|A(t) \rangle $ and $|B(t) \rangle$ such that the 
absolute value of the transition amplitude 
$|\langle B(t)|_Q A(t) \rangle|$ is maximized. 
We proved the theorem by expanding  
$|A(t) \rangle $ and $|B(t) \rangle$ in terms of 
the eigenstates of $\hat{H}$. 
It is noteworthy that, in the future-included CAT 
with a priori non-normal Hamiltonian $\hat{H}$,  
we nevertheless have got a real average for $\hat{\cal O}$ 
at any time $t$ 
by means of the simple expression 
$\langle \hat{\cal O} \rangle_Q^{BA}$.

As for an emerging hermiticity, in ref.\cite{Nagao:2010xu} 
we presented a mechanism to obtain 
a $Q$-Hermitian Hamiltonian 
by considering a long time development. 
The maximization principle studied in this letter 
is another approach to obtaining 
such a $Q$-Hermitian Hamiltonian. 
We have seen that the non-hermiticity of 
the fundamental Hamiltonian $\hat{H}$ has 
disappeared 
from the usually expected results of the model. 
It is this remarkable result of our works with 
non-Hermitian Hamiltonians or complex actions 
that allows us to consider such models to be viable. 
We would not have been able to see any effects 
of the anti-Hermitian part as far as the reality 
of the dynamical variables and the equations of  motion 
are concerned. 
However, as earlier discussed in ref.\cite{Bled2006} 
and also seen in eqs.(\ref{abinotinA0})-(\ref{nc_ATABTB3}), 
the anti-Hermitian part has a strong influence 
on the initial state, which should effectively be seen. 
Indeed, the maximization principle has resulted in a periodicity 
of the history of the universe 
that the initial and final states become basically the same. 
Such an influence would be more recognizable 
in a system defined with a time-dependent non-Hermitian 
Hamiltonian\cite{Fukuma:2013mx}. 
We expect the future-included CAT to 
have the feature that it can provide a unification of an 
initial condition prediction and an equation of motion. 
In this letter, we studied the version defined with 
the $Q$-normalized initial and final states.  
It would be interesting to see what kind of result 
we could obtain in the other version 
defined with the usually normalized initial and final states, 
which is more difficult to study than the 
the version studied here, 
because we cannot fully utilize the 
orthogonality of the eigenstates of the Hamiltonian $\hat{H}$. 
In the future we hope to investigate this version and 
to see if the reality of $\langle \hat{\cal O} \rangle_Q^{BA}$, 
emerging Hermitian Hamiltonian, and such a periodicity 
are suggested or not.

Finally, assuming that the fundamental non-normal Hamiltonian 
$\hat{H}$ is written in terms of Hermitian 
coordinate and momentum operators $\hat{q}$ and $\hat{p}$ 
as $\hat{H}=H(\hat{q}, \hat{p})$, 
we give a possible procedure\footnote{For simplicity, 
we do not use 
the complex coordinate and 
momentum formalism\cite{Nagao:2011za} 
just by supposing the case where the eigenvalues $q$ and $p$ 
are essentially real. 
If we like, we could generalize the argument here 
by following ref.\cite{Nagao:2011za} 
so that we could deal with complex $q$ and $p$.} 
to formulate the $Q$-Hermitian Hamiltonian $\hat{H}_{Qh}$ 
in terms of $Q$-Hermitian 
coordinate and momentum operators $\hat{q}_{Q}$ and $\hat{p}_{Q}$. 
We also introduce a $Q$-Hermitian probability density operator as 
an example of $Q$-Hermitian $\hat{\cal O}$, and construct 
a conserved probability current density. 
Let us begin with defining  
$\hat{q}_{Q}$ and $\hat{p}_{Q}$ by 
\begin{eqnarray}
&&\hat{q}_{Q} \equiv \frac{\hat{q} +  \hat{q}^{\dag^Q} }{2 }, \\
&&\hat{p}_{Q} \equiv \frac{\hat{p} +  \hat{p}^{\dag^Q} }{2 }. 
\end{eqnarray}
Since $Q$ depends on $\hat{q}$ and $\hat{p}$ via $\hat{H}$, 
$\hat{q}_{Q}$ and $\hat{p}_{Q}$ could be written 
in terms of $\hat{q}$ and $\hat{p}$, 
and vice versa.\footnote{In the harmonic oscillator model\cite{KNHBN_ho} defined by the Hamiltonian 
$\hat{H}_{\rm{ho}} \equiv \frac{\hat{p}^2}{2m}+\frac{1}{2}m\omega^2 \hat{q}^2$ 
with a mass $m$ and an angular frequency $\omega$, we obtain 
$\hat{q}_{Q}=  e^{i \frac{\theta}{2}} \hat{q}$ and 
$\hat{p}_{Q}= e^{-i \frac{\theta}{2}} \hat{p}$, where 
$\theta=\arg (m\omega)$. $\hat{H}_{\rm{ho}}$ is rewritten as $\hat{H}_{\rm{ho}}=\frac{\hat{p}_Q^2}{2m_{\rm{eff}}}+\frac{1}{2}m_{\rm{eff}}\omega^2 \hat{q}_Q^2$, where 
$m_{\rm{eff}}=me^{-i\theta}$.} 
Then $\hat{H}$ would be rewritten as 
$\hat{H}=H_{\rm{eff}}(\hat{q}_{Q}, \hat{p}_{Q})$, 
where $H_{\rm{eff}}$ is some analytic function of 
$\hat{q}_{Q}$ and $\hat{p}_{Q}$, 
and $\hat{H}_{Qh}$ is expressed in terms of 
$\hat{q}_{Q}$ and $\hat{p}_{Q}$ as 
\begin{equation}
\hat{H}_{Qh}
=\frac{1}{2} \left( H_{\rm{eff}}(\hat{q}_{Q}, \hat{p}_{Q}) 
+ H_{\rm{eff}}(\hat{q}_{Q}, \hat{p}_{Q})^{\dag^Q} \right) . 
\end{equation}  
Next we define $|q \rangle^Q$ 
as the eigenstate of $\hat{q}_{Q}$ by 
$\hat{q}_{Q} |q \rangle^Q = q|q \rangle^Q$ and 
${}^Q\langle q |_Q q' \rangle^Q = \delta(q-q')$, 
which suggests 
$\int_{-\infty}^\infty dq |q \rangle^Q {}^Q\langle q |_Q =1$. 
Similarly, $|p \rangle^Q$ is introduced as the eigenstate of $\hat{p}_{Q}$ by $\hat{p}_{Q} |p \rangle^Q = p|p \rangle^Q$ and ${}^Q\langle p |_Q p' \rangle^Q = \delta(p-p')$. 
Now, utilizing $|q \rangle^Q$, 
we define the $Q$-Hermitian probability density operator 
\begin{equation}
\hat{\rho} \equiv | q \rangle^Q  {}^Q\langle q |_Q
\end{equation} 
as an example of $Q$-Hermitian $\hat{\cal O}$, 
and write a $q$-representation of  the maximizing state 
$| \tilde{A}(t) \rangle$ as 
\begin{equation}
\psi_{\tilde{A}}(q) \equiv {}^Q\langle q |_Q \tilde{A}(t) \rangle. 
\end{equation}
Then the probability density 
$\rho \equiv \langle \hat{\rho} \rangle_Q^{BA}$ 
is given via the maximization principle by 
$\rho =\langle \hat{\rho} \rangle_Q^{\tilde{A} \tilde{A}} 
= | \psi_{\tilde{A}}(q) |^2$, 
which obeys $\int_{-\infty}^{\infty} dq \rho =1$, 
so  we could construct 
a conserved probability current density 
\begin{equation}
j(q,t) =  \frac{i\hbar}{2m} 
\left(  \frac{\partial \psi_{\tilde{A}}^{*}  }{\partial q}  
\psi_{\tilde{A}} - \psi_{\tilde{A}}^{*}  
\frac{\partial \psi_{\tilde{A}} }{\partial q} \right),
\end{equation}
which satisfies the continuity equation 
$\frac{\partial \rho}{\partial t } +  \frac{\partial }{\partial q} j(q,t) = 0$. 
In realistic cases, not only the maximizing state but also many other states contribute to the transition amplitude, 
while the above relations are obtained 
by considering only the maximizing state, which is 
a kind of approximation in the sense that 
we are ignoring the effects of the other states. 
But we expect that their contribution becomes very small 
in the large $T=T_B-T_A$ case, which we are interested in 
from a phenomenological point of view. 
The larger $T$ we consider, the more the states with the largest positive imaginary part of energy get to dominate.  
Thus we have briefly given 
a possible procedure to formulate 
$\hat{H}_{Qh}$ in terms of $Q$-Hermitian $\hat{q}_Q$ 
and $\hat{p}_Q$, and also constructed 
a conserved probability current density for the maximizing state. 
However, it is not trivial at all 
to determine the local expression of $\hat{H}_{Qh}$ in $q$-space, nor 
to examine the classical behavior of 
$\langle \hat{\cal O} \rangle_Q^{BA}$ explicitly. 
We postpone these problems to future studies.

\section*{Acknowledgements}

K.N. would like to thank the members and visitors of NBI 
for their kind hospitality. 
H.B.N. is grateful to NBI for allowing him to work at the Institute as emeritus.




\begin{thebibliography}{99}


\bibitem{Bled2006}
H.~B.~Nielsen and M.~Ninomiya, 
the proceedings of Bled 2006 
-What Comes Beyond the Standard Models-, 
p.87-124 
(arXiv:hep-ph/0612250). 




\bibitem{Nielsen:2008cm}
  H.~B.~Nielsen and M.~Ninomiya, 
  arXiv:0802.2991 [physics.gen-ph]. 
%


\bibitem{Nielsen:2007ak}
  H.~B.~Nielsen and M.~Ninomiya,
  Int.\ J.\ Mod.\ Phys.\  A {\bf 23}, 919 (2008). 
%




\bibitem{Nielsen:2005ub}
  H.~B.~Nielsen and M.~Ninomiya,
  Prog.\ Theor.\ Phys.\  {\bf 116}, 851 (2007). 



\bibitem{Nielsen:2007mj}
H.~B.~Nielsen and M.~Ninomiya, 
the proceedings of Bled 2007 
-What Comes Beyond the Standard Models-
, p.144-185 (arXiv:0711.3080 [hep-ph]).



\bibitem{newer1}
  H.~B.~Nielsen and M.~Ninomiya,
  arXiv:0910.0359 [hep-ph]. 



\bibitem{Vaxjo2009}
H.~B.~Nielsen, 
%
Found.\ Phys.\ {\bf 41}, 608 (2011). 






\bibitem{newer2}
H.~B.~Nielsen and M.~Ninomiya, 
the proceedings of Bled 2010 -What Comes Beyond the Standard Models-, 
p.138-157(arXiv:1008.0464 [physics.gen-ph]). 

  





\bibitem{Nielsen2010qq}
H.~B.~Nielsen,
arXiv:1006.2455 [physic.gen-ph].


\bibitem{degenerate}
H.~B.~Nielsen and M.~Ninomiya,
arXiv:hep-th/0701018.





\bibitem{Nielsen2009hq}
  H.~B.~Nielsen,
arXiv:0911.3859 [gr-qc].




\bibitem{Bled2010B}
H.~B.~Nielsen, M.~S.~Mankoc~Borstnik, K.~Nagao and G.~Moultaka, 
%
the proceedings of Bled 2010 
-What Comes Beyond the Standard Models-, 
p.211-216 (arXiv:1012.0224 [hep-ph]). 









\bibitem{Bender:1998ke}
  C.~M.~Bender and S.~Boettcher,
  Phys.\ Rev.\ Lett.\  {\bf 80}, 5243 (1998).

\bibitem{Bender:1998gh}
  C.~M.~Bender, S.~Boettcher and P.~Meisinger,
  J.\ Math.\ Phys.\  {\bf 40}, 2201 (1999).






\bibitem{Mostafazadeh_CPT_ip_2002}
A. Mostafazadeh, 
J.\ Math.\ Phys.\ {\bf 43}, 3944 (2002).  
%



\bibitem{Mostafazadeh_CPT_ip_2003}
A. Mostafazadeh, 
J.\ Math.\ Phys.\ {\bf 44}, 974 (2003).








\bibitem{Nagao:2010xu}
K.~Nagao and H.~B.~Nielsen,
Prog.\ Theor.\ Phys. {\bf 125} No. 3, 633 (2011).










\bibitem{Nagao:2011za}
  K.~Nagao and H.~B.~Nielsen,
Prog.\ Theor.\ Phys. {\bf 126} No. 6, 1021 (2011); 
Prog.\ Theor.\ Phys. {\bf 127} No. 6, 1131 (2012) [erratum]. 





\bibitem{Nagao:2011is}
  K.~Nagao and H.~B.~Nielsen, 
   Int.\ J.\ Mod.\ Phys.\ A.{\bf 27}, 1250076 (2012). 



\bibitem{Nagao:2013eda} 
  K.~Nagao and H.~B.~Nielsen,
Prog.\ Theor.\ Exp.\ Phys. (2013) 073A03. 









\bibitem{Nagao:2012mj} 
  K.~Nagao and H.~B.~Nielsen,
  Prog.\ Theor.\ Exp.\ Phys. (2013) 023B04. 


\bibitem{Nagao:2012ye} 
  K.~Nagao and H.~B.~Nielsen, 
the proceedings of Bled 2012: What Comes Beyond the Standard Models, p.86-93 (2012) (arXiv:1211.7269 [quant-ph]).  





		
\bibitem{AAV}
	Y. Aharonov, D. Z. Albert, and L. Vaidman,
	Phys. Rev. Lett. 
	{\bf 60}, 1351 (1988).





\bibitem{review_wv}
	Y. Aharonov, S. Popescu, and J. Tollaksen,
	Physics Today 
	{\bf 63}, 27 (2010). 







\bibitem{Geyer}
F. G. Scholtz, H. B. Geyer and F. J. W. Hahne, Ann. Phys. {\bf 213}, 74 (1992). 







%
\bibitem{Fukuma:2013mx} 
  M.~Fukuma, Y.~Sakatani and S.~Sugishita, 
Phys. Rev. D {\bf 88}, 024041 (2013).



\bibitem{KNHBN_ho}
  K.~Nagao and H.~B.~Nielsen, work in progress. 



\end{thebibliography}
\end{document}